\documentclass{article}

\usepackage{caption}
\usepackage{blkarray}
\usepackage{subcaption}
\usepackage{xfrac}
\usepackage[nolist,nohyperlinks]{acronym}
\usepackage[margin=1.1in]{geometry}
\usepackage{amsmath}

\acrodef{sLM}[sLM]{stochastic lattice model}
\acrodef{ABM}[ABM]{agent-based model}
\acrodef{AD}[AD]{automatic differentiation}
\acrodef{STE}[STE]{straight-through estimator}
\acrodef{GS}[GS]{Gumbel-softmax}
\acrodef{ABC}[ABC]{approximate Bayesian computation}
\acrodef{SI}[SI]{susceptible-infected}
\acrodef{SGD}[SGD]{stochastic gradient descent}
\acrodef{MSD}[MSD]{mean square displacement}
\acrodef{GD}[GD]{gradient descent}
\acrodef{MSE}[MSE]{mean squared error}
\acrodef{MCS}[MCS]{Monte Carlo step}

\title{Fitting Stochastic Lattice Models Using Approximate Gradients}

\author{\textbf{Jan Schering$^2$, Sander Keemink$^{1,*}$, and Johannes Textor$^{2,*}$} \\  \small $^1$ Donders Institute for Brain, Cognition and Behavior, Department of AI, Radboud University, The Netherlands \\ \small $^2$ Institute for Computing and Information Sciences Radboud University Nijmegen, The Netherlands \\ \small $^*$ These authors contributed equally.}

\begin{document}

\maketitle 

\begin{abstract}
Stochastic lattice models (sLMs) are computational tools for simulating spatiotemporal dynamics in physics, computational biology, chemistry, ecology, and other fields. Despite their widespread use, it is challenging to fit sLMs to data, as their likelihood function is commonly intractable and the models non-differentiable. The adjacent field of agent-based modelling (ABM), faced with similar challenges, has recently introduced an approach to approximate gradients in network-controlled ABMs via reparameterization tricks. This approach enables efficient gradient-based optimization with automatic differentiation (AD), which allows for a directed local search of suitable parameters rather than estimation via black-box sampling. In this study, we investigate the feasibility of using similar reparameterization tricks to fit sLMs through backpropagation of approximate gradients. We consider four common scenarios: fitting to single-state transitions, fitting to trajectories, inference of lattice states, and identification of stable lattice configurations. We demonstrate that all tasks can be solved by AD using four example sLMs from sociology, biophysics, image processing, and physical chemistry. Our results show that AD via approximate gradients is a promising method to fit sLMs to data for a wide variety of models and tasks. 
\end{abstract}


\section{Introduction}

\Acp{sLM} are computational tools for simulating spatiotemporal dynamics in both out-of-equilibrium and stable systems  \cite{haselwandter2008renormalization}. They are applied in a wide range of scientific fields including hydrodynamics \cite{chopard2005cellular,rothman2004lattice}, molecular morphogenesis \cite{dab1991lattice, malevanets1997microscopic}, and cell biology \cite{szabo2013cellular, wortel2021local,scianna2021cellular}. Replicating natural phenomena in silico via \acp{sLM} requires calibrating their parameters to data. This task is challenging for three main reasons. First, \acp{sLM} are typically not analytically tractable, which precludes the computation of likelihood functions. Second, \acp{sLM} are stochastic models, which means that black-box optimization methods developed for deterministic functions (e.g., covariance matrix adaptation \cite{hansen1995adaptation}) cannot be easily applied \cite{beyer7behavior}. Finally, \acp{sLM} are discrete models and hence non-differentiable which prevents the computation of gradients.

Motivated by similar challenges, the adjacent field of agent-based modelling has investigated methods of constructing differentiable \acp{ABM} that are amenable to efficient gradient-based optimization via \ac{AD} \cite{andelfinger2023towards,andelfinger2021differentiable,chopra2022differentiable}. An especially promising method is the combination of reparameterization tricks \cite{kingma2013auto} with \acp{STE} to attain approximate gradient estimates from non-differentiable, stochastic models \cite{chopra2022differentiable}. Reparameterization tricks enable differentiating through stochastic models with respect to some parameters of interest. \Acp{STE} approximate the derivatives of discrete operations during backpropagation while upholding the discreteness constraint during simulation \cite{hu2021handling,neftci2019surrogate,bengio2013estimating}. Combining the methods enables approximate end-to-end differentiation of otherwise non-differentiable models.

A potential benefit of this approach over the current state-of-the-art gradient-free Bayesian methods used for parameter inference \cite{alamoudi2023fitmulticell,alamoudi2022massively,wang2022calibration} is that it allows for a local directed search for suitable parameters, rather than estimating entire parameter distributions. This local search is guided by the gradient information, whereas gradient-free methods such as \ac{ABC} \cite{sisson2018handbook,mengersen2013bayesian,sisson2007sequential} rely on black-box sampling. Thus, gradient-based optimization could potentially converge faster and result in better estimates, especially for higher-dimensional models \cite{andelfinger2023towards,andelfinger2021differentiable}.

Current research on gradient-based optimization of \acp{ABM} has largely focused on network-controlled models \cite{chopra2022differentiable,quera2023some,andelfinger2023towards}. This class of ABMs is closely related to \acp{sLM}. Transferring the approach of combining reparameterization tricks and \acp{STE} to the domain of \acp{sLM} could thus hold similar potential and improve the efficiency of model calibration to data.

Depending on the type of model, the goal of calibrating an \ac{sLM} to data can change significantly. In some applications such as morphogenesis \cite{mordvintsev2020growing}, the goal may be to identify a subset of parameters from which certain types of patterns arise. This requires finding parameters with stable-attractor dynamics. However, when studying non-equilibrium processes like the movement of cells in tissues, the goal may instead be to match the summary statistics of the cell movement to a set of target statistics.

In this work, we provide a clear first proof-of-concept of calibrating \acp{sLM} to data through AD. We consider a range of model types and typical calibration tasks and solve them using \ac{AD} via approximate gradients. \textbf{Our contributions} in this work are the following:

\begin{enumerate}
    \item We identify and specify four common types of \ac{sLM} calibration tasks.
    \item We extend the use of reparameterization tricks from network-based ABMs to the \ac{sLM} framework.
    \item We implement four different \acp{sLM} from different domains in a differentiable fashion using PyTorch.
    \item We show the feasibility of AD via approximate gradients for each of the identified fitting tasks by performing AD on models taken from sociology, biophysics, image processing, and physical chemistry.
\end{enumerate}

\section{Related Works}

The simplest strategy for estimating parameters is to apply a grid search. While simple grid searches still see regular application when estimating \ac{sLM} parameters (e.g. in \cite{chang2021supporting,romero2021public}), the method is slow and has issues scaling to many parameters \cite{chopra2022differentiable}. Hence, recent years have seen an influx of different approaches for more efficient calibration.

The currently most prominent class of methods for parameter estimation of \acp{sLM} is likelihood-free Bayesian inference. At the forefront is the \ac{ABC} framework \cite{alamoudi2023fitmulticell,alamoudi2022massively,wang2022calibration}, which constructs a distance measure between the summary statistics of the data and then performs Bayesian inference via Monte Carlo sampling \cite{sisson2018handbook, sisson2007sequential,mengersen2013bayesian}. \ac{ABC} is a powerful tool with several successful applications \cite{durso2021hcv,carr2021estimating,beaumont2002approximate}. More recently, the work in \cite{dyer2022black} discusses the use of alternative neural simulation-based inference methods for \acp{ABM}. These methods enjoy the benefit of being generally more sample-efficient \cite{lueckmann2021benchmarking} but are less amenable to theoretical study due to their black-box nature. Specifically, the impact of model misspecification is known and studied for \ac{ABC} methods \cite{frazier2020model} but less clear for neural simulation-based inference. Early studies suggest a significant deterioration in the performance of the method under model misspecification \cite{cannon2022investigating}. To combat this, the works in \cite{ward2022robust,kelly2023misspecification} suggest misspecification-robust extensions to neural simulation-based inference.

Gradient-based optimization of \acp{ABM} by approximating the gradient has emerged from the hypothesis that directed point estimates of the mode may be beneficial compared to estimating the full distribution. To this end, \cite{andelfinger2021differentiable,andelfinger2023towards} have investigated the smoothing of discrete transformations to approximate the gradient of \acp{ABM}. While delivering promising results, the method scales exponentially due to its handling of conditional branching. Further, the efficient handling of stochasticity has not yet been explored in-depth. As an alternative, reparameterization of discrete distributions with continuous Gumbel-softmax approximations \cite{jang2016categorical} in combination with \acp{STE} has been proposed and investigated for network-controlled \acp{ABM} \cite{quera2023don,chopra2022differentiable}. Early studies on the robustness of Gumbel-softmax approximations for \acp{ABM} indicate that while the method does not guarantee unbiased low-variance gradients \cite{huijben2022review}, it is practically robust enough to enable accurate inference \cite{quera2023some,chopra2022differentiable}. 

Beyond differentiable \acp{ABM}, the use of derivative-free gradient estimators has seen some consideration in the \ac{ABM} optimization literature. The simplest derivative-free method to estimate gradients is finite-diff\-eren\-cing \cite{shi2021numerical}. In \cite{sumata2013comparison}, finite diff\-eren\-cing is used to calibrate an \ac{sLM} of sea ice. However, finite-difference schemes can suffer from an unboundedly high variance when applied to stochastic models \cite{arya2022automatic}. Smoothed infinitesimal perturbation analysis improves on finite differences by estimating the derivative based on conditional expectations \cite{ho2012perturbation}. While a drawback of this method is the need to define hand-derive gradient estimators from analysis of the model, recent works have taken steps towards automating this process akin to \ac{AD} \cite{arya2022automatic}.

In summary, a large body of recent research has been dedicated to efficient calibration methods for non-differentiable and analytically intractable models. Especially of interest are methods that are amenable to gradient-based optimization via \ac{AD}. To this end, the combination of reparameterization tricks and \acp{STE} to attain approximate gradients have shown promising results for network-controlled \acp{ABM}. The application of this approach to \Acp{sLM} has not previously been investigated but promises similar benefits. Thus, we investigate the feasibility of \ac{AD} via approximate gradients for \acp{sLM}.

\section{Approximating the gradient of sLMs}

Differentiability concerning \ac{AD} refers to information being able to flow backward through the computational graph defined by the model in the form
of partial derivatives. Non-differentiability arises from certain types of nodes in the graph not having a well-defined partial derivative and thus blocking
the backward flow of the gradient. \acp{sLM} generally consist of four types of nodes: \textit{continuous deterministic}, \textit{discrete deterministic}, 
\textit{continuous stochastic} and \textit{discrete stochastic} (Fig. 1).

\begin{figure}[h]
     \centering
     \includegraphics[width=0.8\textwidth]{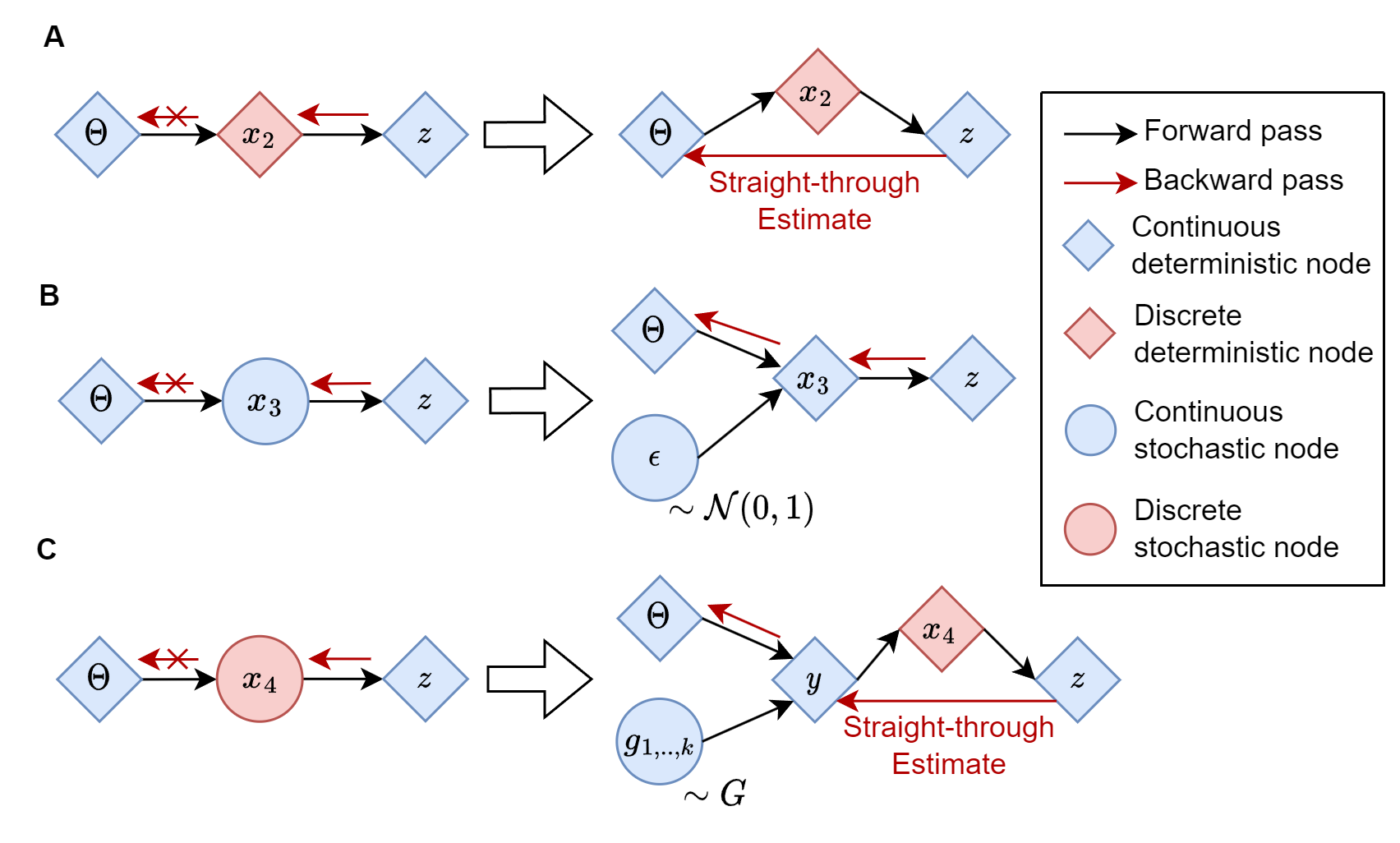}
     \caption{
        \textbf{Gradient flow can be re-enabled with a combination of reparameterization and \acp{STE}.} \\
     \textbf{A} We use an \ac{STE} to estimate the gradient for discrete deterministic nodes.
     \textbf{B} We apply the reparameterization trick to differentiate through continuous stochastic nodes. 
     \textbf{C} The Gumbel-softmax gradient estimator combines reparameterization and \acp{STE} to estimate gradients for stochastic discrete nodes.
     }
     \label{fig:approach_comp}
\end{figure}

Continuous deterministic nodes are differentiable, while the other three types block the backward flow of the gradient as shown on the left side in Fig. \ref{fig:approach_comp}. \acp{sLM} consist overwhelmingly of continuous nodes. That is, we can assume \acp{sLM} to be almost differentiable everywhere with respect to its parameters. The challenge is to find a way to enable the backward flow of gradients through the blocking nodes.

\subsection{Dealing with discreteness}
\label{meth:dis}

\Acp{sLM} commonly make use of discrete transformations in order to determine the new state of a given lattice site. An example is the Heaviside step function $h(x)$,  which maps real numbers onto a binary state space. The derivative of $h(x)$ is the Dirac delta and is 0 everywhere except for the step threshold, where it is infinite. Similarly, the derivative of every discrete transformation is 0 everywhere except for at the thresholds between categories. As a result, information cannot flow backward because no information is passed through the derivative. 

STEs \cite{bengio2013estimating} bypass the ill-defined partial derivative of the discrete transformations by replacing it during the backward pass with an approximate derivative function (the STE). The simplest STE is the identity function. The viability of identity STEs has been shown in the context of spiking neuron models \cite{tang2022relaxation,neftci2019surrogate}. Thus, we choose to employ identity STEs to approximate the gradient for discrete nodes. The identity \ac{STE} approach is illustrated in Fig. \ref{fig:approach_comp}\textbf{A}.

\subsection{Dealing with stochasticity}
\label{meth:stoc}

Stochastic nodes represent the sampling of variables from a distribution. In \acp{sLM}, the shape of the distributions usually depends on some subset of the model parameters $\Theta$. However, sampling operations do not have a well-defined derivative. Hence, we cannot determine the gradient with respect to the parameters that determine the shape of the distribution. 

The reparameterization trick \cite{kingma2013auto} removes the dependency of the sampling operation on the model parameters $\Theta$. To achieve this, an auxiliary random variable $\epsilon$ distributed by some distribution $p(\epsilon)$ that does not depend on the model parameters $\Theta$ is introduced. Then, we rewrite $z = g(\Theta, X_t, \epsilon)$ into a deterministic transformation $g$ of the parameters, the lattice state, and the auxiliary variable. The result is equivalent to drawing a sample from the original distribution. Rewritten in this way, gradient information can flow independently of the random sampling operation. An illustration of this is shown in Fig. \ref{fig:approach_comp}\textbf{B}.

\subsection{The Gumbel-softmax trick}
\label{meth:GS}

The \ac{GS} trick \cite{jang2016categorical,maddison2014sampling} combines reparameterization with \acp{STE} in order to estimate the gradient for the special case of discrete stochastic nodes. For reparameterization, auxiliary noise variables $g_{1,2,\dots,n}$ are sampled from the Gumbel distribution $G$ and added to the log-probability of each class $\alpha_{1,2,\dots,n}$. Then, softmaxing is applied. The result is a continuous approximation of the original categorical distribution. In order to uphold the discreteness of the model, the samples are further discretized to one-hot vectors using the arg max function. To enable gradient flow during the backward pass, an identity \ac{STE} replaces the derivative of the arg max function. The \ac{GS} gradient estimator approach is illustrated in Fig. \ref{fig:approach_comp}\textbf{C}.

\subsection{Implementing the approach in software}

Each model was implemented using the PyTorch library \cite{stevens2020deep} version 2.0.1. PyTorch is a modular Python framework that can be applied to optimize algebraic models by providing automatic differentiation tools \cite{paszke2017automatic}. The code for all experiments is available from the authors upon request.

\section{Defining sLM Calibration Tasks}

\ac{sLM} are used in many fields and are capable of simulating a broad range of phenomena. These phenomena can have significantly varying aspects of interest like stable-attractor dynamics for morphogenesis models or out-of-equilibrium dynamics in swarm models. 
Calibration of \acp{sLM} to data thus defines a class of parameter fitting tasks, the applicability of which depends on the type of model, the data available, and the target behavior to emulate. It is then beneficial if a calibration method is generally applicable to at least the most common types of tasks. In this work, we consider the following four types of fitting tasks, which cover most of the applications of sLMs that we are aware of; however, this is not intended to be an exhaustive list, and it is possible that there are further relevant fitting tasks that do not fit into one of the below categories. 

\textbf{Transition fitting}
is the task of finding parameters, such that the per-timestep behavior of the model matches a set of observed transitions. This type of task applies, whenever significant information can be inferred from the transitional behavior of the model. An example of this is epidemiology. Here, each state transition (e.g. the per-week spread of Covid within a given region) informs our understanding of the spreading behavior. When emulating such a system, it is important to simulate the transitional behavior with high fidelity for the model to be useful in studying the spread of the disease, whereas considering longer-term trajectories for infectious diseases does not necessarily lead to an increase in predictability and can even have adverse effects \cite{scarpino2019predictability}.

\textbf{Trajectory fitting} focuses on the longer-term behavior of the model. It involves finding parameters such that certain summary statistics of trajectories generated by the model match a set of target statistics. Being able to perform trajectory fitting is crucial for studying real-world phenomena of non-equilibrium systems. Most commonly, when modelling real-world phenomena, the main interest lies in the long-term behavior of the system which transition fitting does not adequately capture. Due to the stochasticity of the models, reproducing full trajectories is unlikely. The goal, instead, is to match a set of relevant summary statistics to a given target. Example applications of this are systems that exhibit Brownian motion such as stellar bodies \cite{merritt2013dynamics} or living cells \cite{tsekov2013brownian} where one may aim to match summary statistics such as the \ac{MSD} over time.

\textbf{State fitting} is the process of optimizing a lattice configuration with respect to some objective measure. The primary goal of this task is to accurately reflect the state of a real-world system through the configuration of the lattice. Considering for example a \ac{sLM} of forest fire growth such as described in \cite{boychuk2009stochastic}, a potential task is inferring lattice configurations from imaging of real-world forest fires. Another area of interest is cell segmentation where, given some reference cell imaging, the goal is to discriminate between cell pixels and non-cell pixels.

\textbf{Stability fitting} is the task of finding parameter configurations for \acp{sLM} from which stable patterns arise. For instance, 
the formation of such stable patterns from an unstable initial configuration over time is an important aspect of biological morphogenesis. Starting with Turing in 1952 \cite{turing1990chemical}, several reaction-diffusion models for the emergence of stable 'Turing' patterns have been proposed and studied \cite{ali2023spatiotemporal,li2015turing,wang2011numerical,lengyel1992chemical}. Independent of the model choice, Turing patterns have been shown to only emerge for a small subset of the parameter space which is generally hard to identify \cite{scholes2019comprehensive,vittadello2021turing}.

\section{Experiments}

\subsection{Transition fitting of a 1-parameter susceptible-infected sLM}

Transition fitting is the task of fitting a model to a given set of state transitions -- hence, we consider only a single time step of the model. As an example, we use a simple \ac{SI} \ac{sLM} that simulates diffusive spread of a news item through a population. The \ac{SI} \ac{sLM} depends on a single ``spread coefficient'' $\beta$ that determines the probability of the news item travelling from a lattice site to a given neighbour. 
In our experiment, we generate state transition data using known $\beta$ values, and apply AD to recover $\beta$ from the dataset. 

\subsubsection{Model description}

The SI model used here is a spatial variation of a Markov-chain model described in \cite{conlisk1976interactive}. In this spatial variation, the \ac{sLM} is defined as a binary 2D square lattice on a periodic torus. Lattice sites in state 1 are considered `aware' of the news while sites in state 0 are `unaware'. Denoting $L_t^{(i,j)}$ as the state of the lattice at site $(i,j)$ at time $t$ and $N^{(i,j)}$ as the number of aware sites in the Moore neighborhood of $L^{(i,j)}$, the update rule is:

\[ 
p\left(L^{(i,j)}_{t+1} \mid L^{(i,j)}_t,N^{(i,j)}_t,\beta\right) =
\begin{blockarray}{cc|c}
1 & 0 &  \sfrac{L^{(i,j)}_t}{L^{(i,j)}_{t+1}}\\ \hline
\begin{block}{cc|c}
1 & 1 - (1-\beta)^{N^{(i,j)}_t} & 1 \\
0 & (1-\beta)^{N^{(i,j)}_t}     & 0 \\
\end{block}
\end{blockarray}
\]

The first column of the probability matrix encodes that news items cannot be forgotten once learned. The second column models the likelihood of a lattice site turning aware or staying unaware, depending on the number of aware neighbors. At each time step, this rule is applied synchronously to each lattice site. That is, the new value assigned to each lattice site at a time $t+1$ is sampled from the discrete binary distribution $p(L^{(i,j)}_{t+1}|L^{(i,j)}_t,N^{(i,j)}_t,\beta)$. 

The differentiation problem in this case is that the probability $p(L^{(i,j)}_{t+1}|L^{(i,j)}_t,N^{(i,j)}_t,\beta)$ is discrete and the sampling depends on the model parameter $\beta$. Hence, we cannot differentiate through the model with respect to $\beta$. We apply the \ac{GS} gradient estimator, as described in Sec. \ref{meth:GS}, to circumvent this issue (Fig. \ref{fig:approach_comp}C). 

\subsubsection{Model training}

We define target values for $\beta$ in range $[0,1]$. For each target value, a dataset of state transitions $(X_t, X_{t+1})$ is collected by running $N$ reference simulations with the target value. Each reference simulation is initialized with a single aware site in the center of the lattice and run for $\tau$ steps. For $\beta \in [0.05, 0.5]$ we use $\tau=50$ and $N=100$. For $\beta > 0.5$, we adjust to $\tau=30$ and $N=200$ as the rate of spread reaches the boundaries of the grid faster. 

For gradient optimization, minibatch \ac{SGD} \cite{li2014efficient} is applied with a batch size of 128. Further, the pixel-wise mean-squared Error (MSE) between predictions $\hat{X}_{t+1}$ and observations $X_{t+1}$ is used as the optimization objective. A learning rate of $1e^{-7}$ was empirically chosen. 

\begin{figure}[h]
    \centering
    \includegraphics[width=0.8\textwidth]{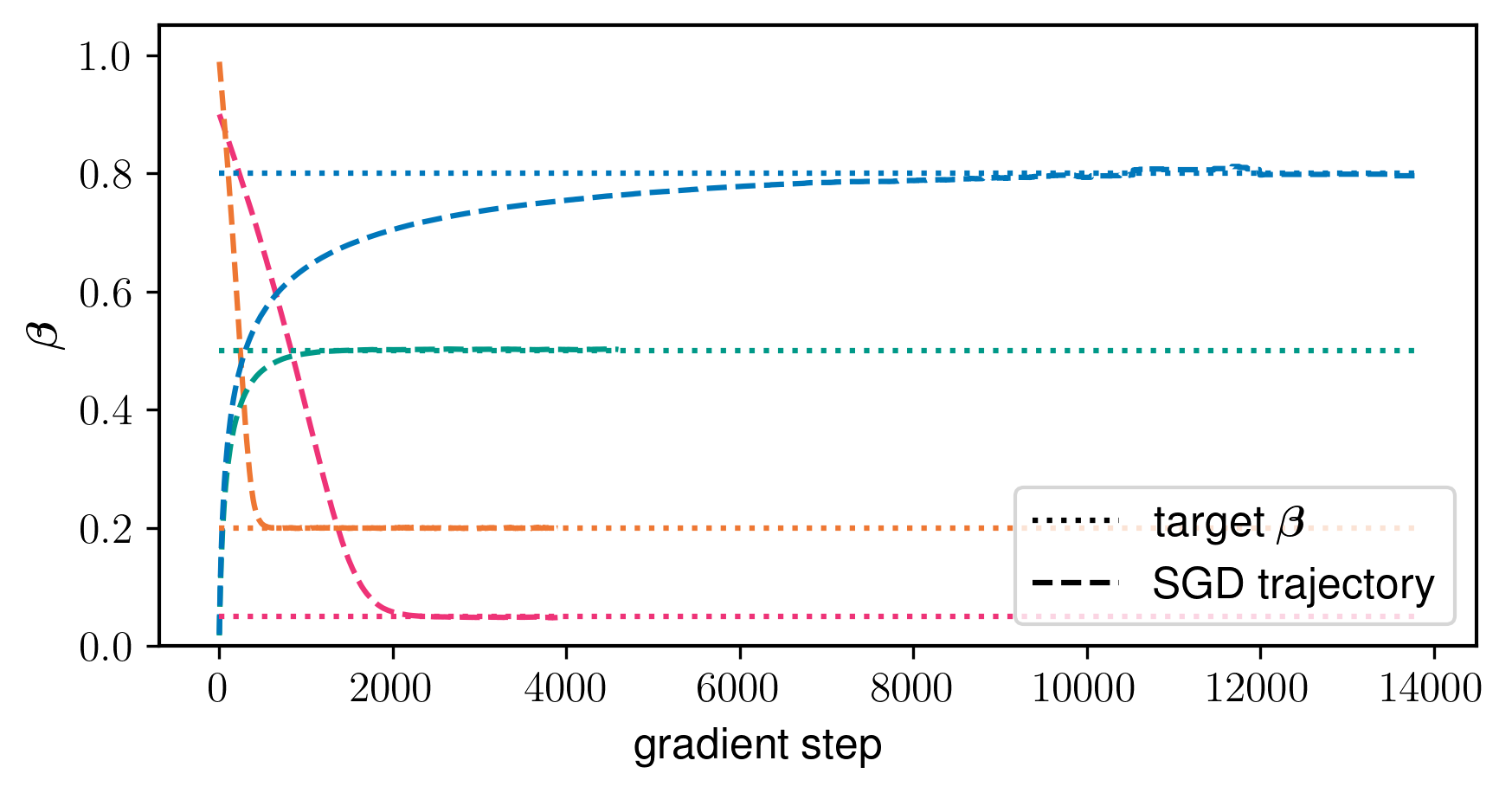}
    \caption{
       \textbf{Transition fitting the spread coefficient of a simple SI \ac{sLM} via AD can recover a close estimate of the original $\beta$ for a range of target values.}
       \\ Each color refers to the results of fitting $\beta$ to a given target value (dotted lines) using SGD via approximate gradients. The dashed lines show the estimate of $\beta$ throughout the optimization process.
    }
    \label{res:si}
\end{figure}

\subsubsection{Training Results}

Fig. \ref{res:si} shows the results of fitting the \ac{SI} \ac{sLM} to a range of different target values for $\beta$. For each of the target values, a closely matching estimate of $\beta$ is successfully recovered. This shows that AD via approximate gradients can successfully be used to perform transition fitting for \acp{sLM}. 

Standing out from the results is that fitting $\beta$ converges significantly slower for higher values of $\beta$ compared to lower values. The reason for this is that for higher spread coefficients the rate of spread saturates and trajectories become very similar to another.

\subsection{Trajectory fitting of a persistent random walk sLM}
\label{sec:cpmr}

Although transition fitting is a straightforward task that is simple to define, in many \ac{sLM} applications we will be interested in capturing the longer-term behaviour of a system. A natural generalization of transition fitting is to consider longer simulation \emph{trajectories} consisting of multiple transitions. Due to the stochasticity of \acp{sLM}, our goal will generally not be to exactly reproduce a given set of trajectories; instead, we will normally want to match some given set of summary statistics between our model and the data. 

As an example, we consider a persistent random walk \ac{sLM}. For sufficiently many steps, it is highly unlikely that we will exactly match any given random walk sequence, nor would this be a reasonable goal. Instead, random walk models are typically fitted to summary statistics such as the \ac{MSD} over time of a given set of trajectories. Here, we consider the movement of different types of cells that are known to exhibit a persistent random walk over time. Specifically, we apply \ac{AD} via approximate gradients to fit the parameters of a persistent random walk \ac{sLM} to the \ac{MSD} collected in \cite{wortel2021celltrackr} for T cells, B cells, and Neutrophils.

\subsubsection{Model Description}
We model a persistent random walk on a binary regular square lattice as two independent random walks along the $x$ and $y$ axes. For each axis, the current velocity $v_x,v_y \in \{-1,0,1\}$ can be either forward, backward, or staying in place. We define one \ac{MCS} as the combination of taking a step in both $x$ and $y$ with the respective velocity. At every \ac{MCS}, we retain the velocity of the previous step with a probability $1-p_{\text{resample}}$. Thus with a probability of $p_{\text{resample}}$, we resample the velocities according to a distribution:

\[
v = 
\begin{cases}
-1, \quad \text{with } p=\sfrac{(1-p_{\text{center}})}{2}
\\
0, \quad \text{with } p=p_{\text{center}}
\\
1, \quad \text{with } p=\sfrac{(1-p_{\text{center}})}{2}
\end{cases}
\]

It follows that $p_{\text{center}}$ defines the probability of staying in place for the $x$ and $y$ direction respectively.
Both $p_{\text{center}}$ and  $p_{\text{resample}}$ are logit-transformed during fitting for improved numerical stability.

The model contains two discrete sampling operations. First, the binary decision of resampling or keeping the current velocity depending on the model parameter $p_{\text{resample}}$. Second, sampling a new velocity from the categorical distribution over $\{-1,0,1\}$. To make the model differentiable, we approximate the gradient of the sampling operations using the \ac{GS} gradient estimator as shown in Fig. \ref{fig:approach_comp}C and described in Sec. \ref{meth:GS}.

\subsubsection{Model training}

As a loss function, we calculate the sum of \acp{MSE} between the \ac{MSD} of the simulated batch and the target \ac{MSD} for every simulation step. We apply the Adam optimization scheme \cite{kingma2014adam} with an empirically chosen learning rate of 0.01. For each step of \ac{GD}, we simulate a batch of 1,500 lattices for 10 steps to match the number of datapoints. 

For recovering parameters from synthetically generated data, we simulate a batch of 15,000 trajectories using
$p_{\text{center}}=0.2,p_{\text{resample}}=0.1$. We track the \ac{MSD} of the batch over 10\ac{MCS} and use these summary statistics as the fitting target. For the cell data, we consider a dataset of trajectories where the \ac{MSD} is tracked periodically every 24 seconds over a certain time frame of which we consider the first ten datapoints. For the persistent random walk \ac{sLM} of the cell data, we define $1\text{MCS} = 24s$ as well as $\Delta x = \sqrt{10}\mu m$. Limiting the datasets to a comparatively small amount of 10 datapoints pronounces the persistence effects, which vanish for larger time frames \cite{furth1920brownsche}.

\subsubsection{Results}

\begin{figure}[h!]
    \centering
    \includegraphics[width=\textwidth]{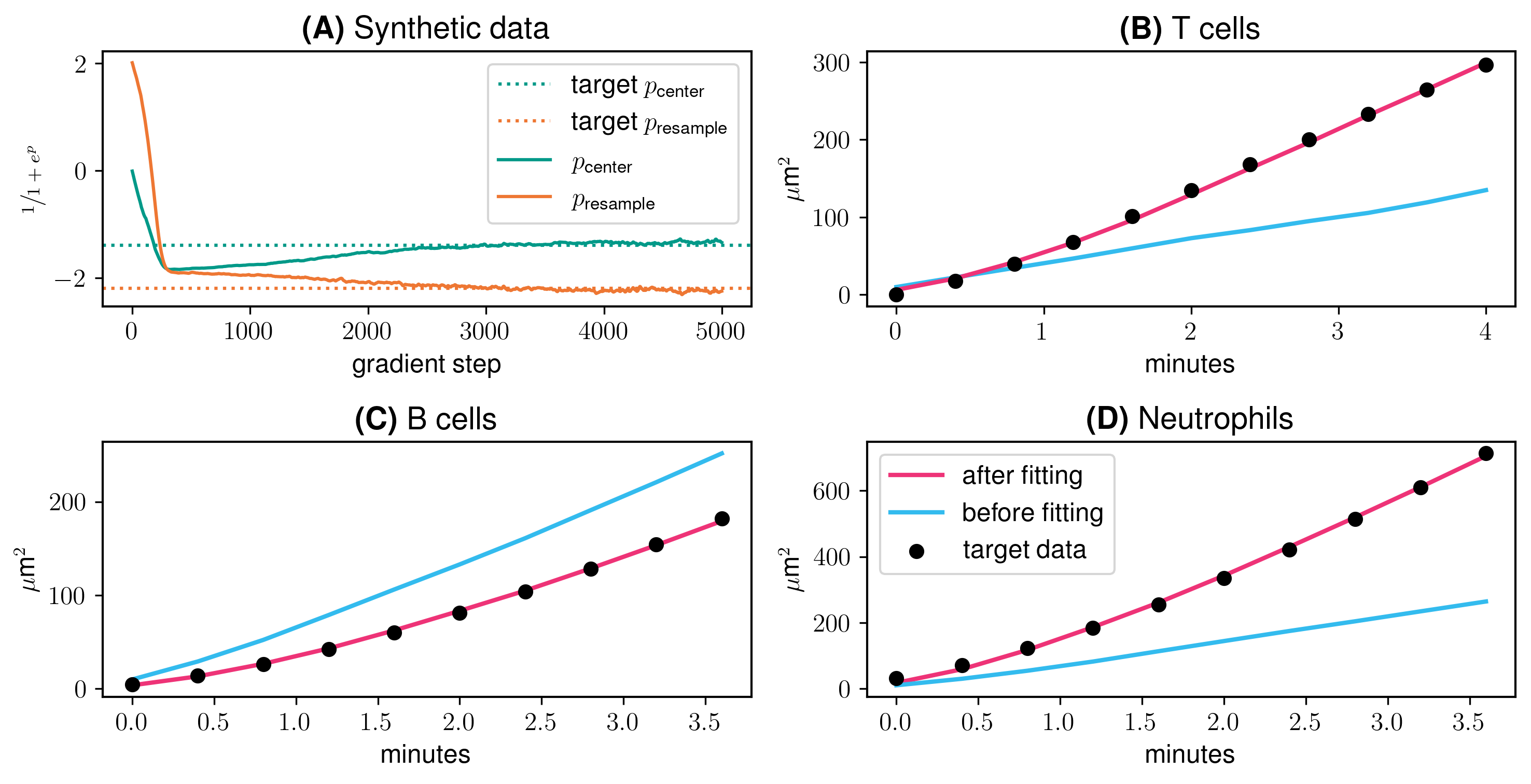}
    \caption{
       \textbf{AD via approximate gradients can successfully perform trajectory fitting on a persistent random walk sLM to datasets of cell movement as well as recover parameters.} \\
    \textbf{(A)} We show that AD via approximate can recover parameters by applying the method to synthetic data.
    \textbf{(B),(C),(D)} We apply trajectory fitting using the MSD curve of (A) a set of T cells, (B) a set of B cells (C) a set of Neutrophils over time.}
    \label{res:cpm_r}
\end{figure}

Fig. \ref{res:cpm_r}A shows the results of recovering the target parameters from synthetically generated data. Starting from bad initial guesses, the estimated parameters converge to closely match the target parameters after about 5,000 steps of gradient descent.

Fig. \ref{res:cpm_r}B shows the results of fitting the persistent random walk \ac{sLM} to a dataset of T cell movement. Persistent brownian motion exhibits a pattern of an initial non-linear `onramping' phase that transitions into quasi-linear growth over time. The steepness of the onramping is mainly governed by the resampling probability $p_{\text{resample}}$ whereas the transition time between onramping and linear growth is governed by the probability of moving $p_{\text{center}}$. After 10,000 steps of Adam optimization, the MSD curve of the batch simulated with the optimized model closely matches the MSD curve and persistence pattern of the T cell dataset.

Fig. \ref{res:cpm_r}C,D show similar results for the experiments using B cell data and Neutrophil data, respectively. In each case, the data follows the persistent random walk pattern of non-linear onramping transitioning into quasi-linear growth of the MSD. We start from bad initial guesses for the parameters, which produce MSD curves that match neither the persistence pattern nor the average displacement over time of the data. After 10,000 steps of Adam optimization in both cases, we attain parameter estimates that produce MSD curves closely matching those of the target data.

The results show that starting from bad initial parameters, we can successfully make use of AD via approximate gradients to estimate a set of parameters $p_{\text{center}},p_{\text{resample}}$ for which the MSD curve closely matches that of the target data. Hence, the results suggest that AD via approximate gradients can be used for trajectory fitting \acp{sLM} to data.

\subsection{State fitting of lattice configurations from images}

Having considered fitting out-of-equilibrium systems to short- and longer-time dynamics, we now turn our attention to fitting tasks that focus on a stable lattice state. First, we will consider directly fitting the lattice state to a given image. An example of such a task is the Potts model for image segmentation \cite{storath2015joint}. Here, each lattice site is a latent variable representing the identity of a class (region or object), and the goal is to find a configuration of lattice sites that fits the given image while respecting a constraint that favours adjacent sites to be of the same class. For simplicity, we consider the two-class Potts model applied to foreground-background segmentation. This could also be used to infer the initial state of an \ac{sLM} from a given reference image of the system to be emulated. 

Foreground-background segmentation can be challenging due to illumination heterogeneity and other issues \cite{long2018foreground}. Recent years have seen an influx of machine learning methods based on customized loss functions \cite{lim2020learning, long2018foreground, wang2017interactive}. Being able to directly optimize grid configurations using AD via approximate gradients implies that we can use the same optimization strategy for initializing \acp{sLM} and calibrating their behavior. 

\subsubsection{Model description}

We consider a foreground-background segmentation model that maps from a given $n \times m$ RGB image $X \in  [0,255]^{n \times m \times 3}$ to a binary lattice mask $X_{\text{pred}} \in \{0,1\}^{n \times m}$. Each lattice site $X_{\text{pred}}^{(i,j)}$ predicts whether the corresponding pixel $X^{(i,j)}$ belongs to the foreground (state 1) or the background (state 0).

To this end, we first transform the image to greyscale and normalize it into the range $[0,1]$. We define the resulting greyscale image as $X_g$. Further, we store a copy $X_{\text{ref}} := X_g$ of $X_g$ for use during loss calculations. To discretize $X_g$ into a binary lattice configuration $X_{\text{pred}}$, we apply a threshold function:

\[
X^{(i,j)}_{\text{pred}} = 
\begin{cases}
1, \quad \text{if } X_g > 0
\\
0, \quad \text{otherwise}
\end{cases}
\]

to every pixel $(i,j)$ of the greyscale image. Hence, pixels with an intensity above zero are classified as foreground, while pixels with an intensity of exactly zero are classified as background. The goal is then to adjust the pixel intensities of $X_g$, such that the discretized lattice $X_{\text{pred}}$ yields the best segmentation performance.

The problem with applying \ac{AD} to optimize the lattice $X_{\text{pred}}$ directly is that it is discrete. Hence, the derivative of the segmentation loss with respect to the binary lattice sites would be ill-defined.
The key to solve this is that we calculate the loss based on the greyscale image $X_g$ itself rather than the binary lattice, disregarding the impact of the discretization step. Hence, we implicitly define a Straight-through estimate of the true loss and gradient, as described in Sec. \ref{meth:dis} and shown in Fig. \ref{fig:approach_comp}A.

\subsubsection{Model training}

A custom loss function was defined to capture this objective:

\begin{itemize}
    \item $l_a = \sum_{(i,j)} 1 - X^{(i,j)}_g$ penalizes pixels that approach 0.
    \item $l_b = \sum_{(i,j)}\sum_{u \in N(x)} X^{(i,j)}(1 - u)$ penalizes high-intensity pixels surrounded by low intensity pixels. The underlying assumption is that foreground pixels usually have contact with multiple other foreground pixels, e.g. forming objects.
    \item $l_c = \sum_{(i,j)}\sum_{u \in N(x)} (1-X^{(i,j)})u$ penalizes low-intensity pixels surrounded by high-intensity pixels, similar to $l_b$.
    \item $l_d = \sum_{(i,j)} (X_g - X_{\text{ref}})^2$ ensures that important information in the image is retained.
\end{itemize}

\noindent $N(x)$ is the local Moore neighborhood of pixel $x$ while $Coords(X)$ is a function that returns a list of all pixel coordinates $(i,j)$ for the image. The total loss function can then be written as $\mathcal{L}(X_{g}) = \lambda_a l_a + \lambda_b l_b + \lambda_c l_c + \lambda_d l_d$. The $\lambda$ terms are scaling factors of the different components, allowing us to weigh their relative importance. With the loss function in place, $X_{g}$ is optimized towards $\mathcal{L}$ using GD. 

\subsubsection{Results}
Segmentation performance is shown for two sample images from separate domains. First, the method is applied to a natural input image taken from the DIS5K dataset introduced in \cite{qin2022highly}. Then, the method is applied to a sample cell image taken from the BBBC005v1 Broad Bioimage Benchmark Collection introduced in \cite{ljosa2012annotated}. For the DIS5K seadragon image, GD was applied to minimize $\mathcal{L}(X_{pred})$ with a learning rate of $0.01$, $\lambda_a = 0.6$, $\lambda_b=0.05$, $\lambda_c=0.15$ and $\lambda_d = 0.5$. For the BBBC005v1 cell imaging, GD was applied with the same learning rate and $\lambda_a = 0.6$, $\lambda_b=0.05$, $\lambda_c=0.3$ and $\lambda_d = 0.8$.

\begin{figure}[h]
    \centering
    \includegraphics[width=0.8\textwidth]{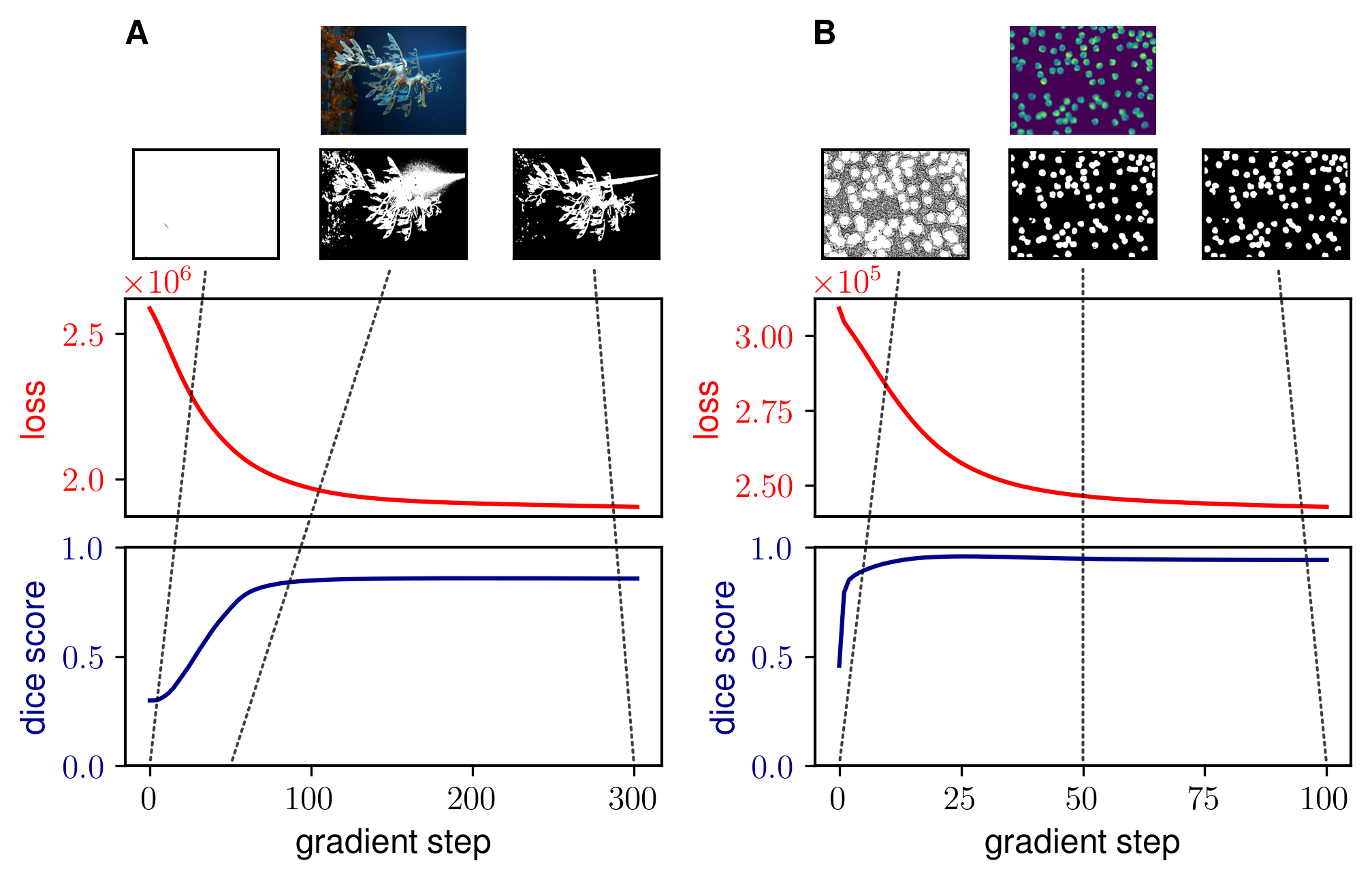}
    \caption{\textbf{State fitting using AD via approximate gradients achieves promising results for sample images from two separate domains.} \\
    \textbf{A} State fitting applied to a natural image of a seadragon.
    \textbf{B} State fitting applied to cell imaging. 
    \textbf{Top Row 1} The reference image. 
    \textbf{Top Row 2} Sample lattices throughout the fitting process.
    \textbf{Middle Row} Loss (red) per experiment over time.
    \textbf{Bottom Row} Dice score (blue) per experiment over time.
    }
    \label{res:segment}
\end{figure}

The segmentation performance for both images is illustrated in Fig. \ref{res:segment}. We report segmentation performance via \textit{dice score} between predictions and ground-truth segmentation masks. In addition to the dice score, the loss per gradient step is shown for each image. Finally, samples of the segmentation mask at different points of the fitting process are shown above the plots.

Fig. \ref{res:segment}A illustrates segmentation performance for experiment 1. The fitting process starts from a very low initial dice score of $\sim0.30$ indicating that no segmentation is happening. The initial foreground mask confirms this, showing that initially every pixel is categorized as foreground (white). Within the first 50 gradient steps, segmentation performance improves close to linearly to a score of  $\sim0.8$. This implies a significant overlap of the predicted segmentation with the ground truth. The corresponding segmentation mask confirms this, clearly showing the shape of the seadragon. The loss converges after about 300 steps of gradient descent, where we achieve a dice score of $0.85$. The corresponding foreground mask captures the seadragon quite well. The remaining misclassifications stem from some bright background artifacts and some dark spots in the skin pattern of the seadragon. 

Fig. \ref{res:segment}B illustrates segmentation performance for a sample image from the BBBC005v1 cell imaging dataset. Initial segmentation performance is a bit higher than for the seadragon but still very low with a dice score of $0.45$. In the initial foreground masks, cells are visible but the mask is very noisy and the predicted cell segments are overly large. After 20 steps of gradient descent, the noise has been successfully removed and cell sizes in the foreground mask match those of the input image. The dice score here is 0.95, indicating an almost perfect match with the ground truth. Over the next 80 steps of gradient descent, the loss does decrease further, while the dice score instead decreases slightly and plateaus at 0.94. A possible reason for the slight decrease in dice score despite the loss going down, is that the loss function does not fully capture the segmentation objective. Thus, in this case, overfitting to the loss function comes at a loss of segmentation performance.

Overall, the results of this experiment suggest that \ac{AD} using approximate gradients can in principle be used to fit the configuration of \ac{sLM} lattices. Segmentation performance for both samples is high, with a final score 0f 0.85 for the seadragon and 0.94 for the cell image. The loss function can be expanded and adjusted to incorporate more domain knowledge, possibly increasing the performance further.

\subsection{Stability fitting of a reaction-diffusion sLM for pattern generation}

Finally, we investigate applying stability fitting to a reaction-dif\-fu\-sion \ac{sLM} that produces stable patterns for certain regions of the parameter space. Being able to apply AD for this type of fitting would provide an efficient way of locating the subspaces that generate Turing patterns. Further, the training process can provide additional insights into the model, highlighting the impact of the different parameters on pattern formation.  

\subsubsection{Model description}

We consider the ``Malevanets-Kapral" reaction-diffusion \ac{sLM} introduced in \cite{malevanets1997microscopic}. The Malevanets-Kapral model is a spatial reformulation of the Fitzhugh-Nagumo equations \cite{fitzhugh1961impulses}, which simulates the reaction-diffusion of two chemical species over time. For an in-depth description of the model we refer to \cite{malevanets1997microscopic}. Generally, the Malevanets-Kapral sLM models per-chemical concentration of two chemical species A, B in space as a 2-layer square lattice $L=(A,B)$. $A^{(i,j)}$ represents the concentration of species A at the lattice site $(i,j)$, whereas $B^{(i,j)}$ denotes the concentration of species B at the same lattice site. Every site has a maximum capacity of $N$ molecules per species. At every time step, the model simulates:

\begin{enumerate}
    \item Independent diffusion of $A$ and $B$, governed by the diffusion coefficients $D_A,D_B$
    \item The reaction of A and B with each other, using mass-action kinetics with reaction rates $k_1,k_2,k_3$
\end{enumerate}

Importantly for our experiments, the non-differentiability of this model stems from two sources. First, at every lattice site a reaction happens with a probability depending on the reaction rate of the corresponding reaction channel. We approximate the gradient of this sampling operation by applying reparameterization as described in Sec. \ref{meth:stoc} in combination with a \ac{STE} (Fig. \ref{fig:approach_comp}A). Second, diffusion involves random direction sampling followed by translating the grid in the chosen direction. To approximate the gradient of the categorical sampling, we apply a \ac{GS} gradient estimator (Sec. \ref{meth:GS}, Fig. \ref{fig:approach_comp}C).

Additionally, a parameter $\gamma$ is used as a timescale of the reaction process in relation to the diffusion process. By iteratively applying the two steps to the lattice, the reaction-diffusion process of the two chemicals is simulated. For a subset of parameters, this interaction generates stable labyrinthine patterns, as shown in the top row of Fig. \ref{res:fhn}. 
Here, we test: 

\begin{itemize}
    \item [\textit{E1}]: Joint optimization of the model parameters \\ $\Theta = (D_A,D_B,k_1,k_2,k_3)$
    \item [\textit{E2}]: Optimization of the reaction rates $k_1,k_2,k_3$ for fixed diffusion coefficients
    \item [\textit{E3}]: Optimization of the diffusion coefficients $D_A,D_B$ for fixed reaction rates
\end{itemize}

We consider these three different experiments to investigate whether reaction or diffusion rates are more difficult to find, and whether joint optimization of both is feasible.

\subsubsection{Model training}
Pattern formation performance is challenging to accurately capture in a loss function \cite{vittadello2021turing}. Inspired by the training strategy used by \cite{mordvintsev2020growing} for optimizing continuous 'Neural Cellular Automata' models towards generating stable patterns, we use pattern maintenance as an alternative measure that is easily quantifiable. Thus for each gradient step, the evolution of a given stable pattern $X_{ref}$ is simulated for $\tau$ timesteps to produce $X_{ref+\tau}$. The pixel-wise MSE of $(X_{ref}, X_{ref+\tau})$ then serves as a surrogate optimization objective. To validate how well this generalizes to pattern formation, we perform \textit{pattern formation tests} during optimization using the current parameters. In each test, we simulate a system starting from a uniform, unstable state and observe if stable patterns emerge.

Each experiment is performed on $64\times64$ regular grid lattices on a periodic torus. For optimization, standard gradient descent is applied with an empirical learning rate of $0.05$. Each simulation is run for $\tau=500$ steps with a reaction timescale $\gamma=0.005$ and maximum capacity $N=50$. For experiment 2, we fixed $k_1=0.98$, $k_2=0.1$ and $k_3=0.2$, based on results reported in \cite{malevanets1997microscopic}. Similarly, for experiment 3 we fixed $D_A=0.1$ and $D_B=0.4$. 

\subsubsection{Training Results}

\begin{figure*}[h!]
    \centering
    \includegraphics[width=\textwidth]{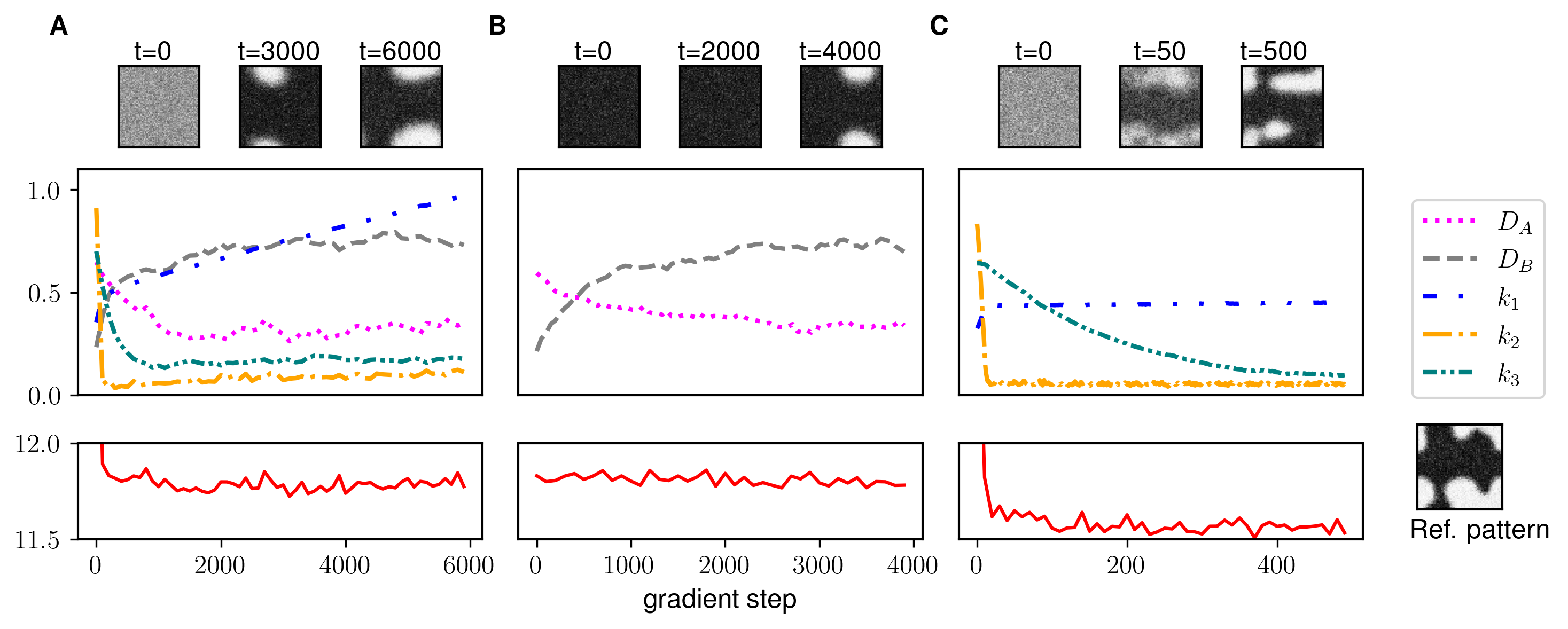}
    \caption{\textbf{Morphogenetic parameter configurations of the Malevanets-Kapral model are found through gradient-based stability fitting using AD.}\\
    \textbf{A} Joint fitting of the reaction rates and diffusion coefficients. 
    \textbf{B} Fitting of the diffusion coefficients for fixed reaction rates. 
    \textbf{C} Stability fitting applied to the reaction rates, with fixed diffusion coefficients.
    \textbf{Top Row} samples of the pattern formation test taken at different points of the fitting process for each experiment. The samples show the concentration of the A species after fifty-thousand simulation steps, starting from a uniform distribution of A and B chemicals. 
    \textbf{Middle Row} Parameter trace throughout the fitting process. 
    \textbf{Bottom Row} Pixel-wise square difference throughout the fitting process. 
    }
    \label{res:fhn}
\end{figure*}

For experiment \textit{E1}, 6,000 steps of gradient descent are shown in Fig. \ref{res:fhn}A. For the initial parameters, no patterns are formed. After 3,000 steps, the test sample shows an overall higher concentration of species A with a small pattern in the spatial distribution. Finally, the test sample after 6,000 steps shows clear patterns in the spatial distribution of species A. Thus, the gradient-based fitting method has successfully identified a set of morphogenetic parameters.
The loss curve for experiment 1 shows that the loss drops significantly within the first few steps of gradient descent. Looking at the parameter trace, this seems to be largely driven by the reaction rate $k_2$.

The results of experiment \textit{E2} are summarized in Fig. \ref{res:fhn}B. When comparing the pattern samples to the test samples of experiment E1, some interesting details stand out. Foremost, the initial parameter configuration does not produce a white noise pattern. Instead, the concentration of species A appears to have converged to maximum capacity for all lattice sites. This state is stable, but no patterns are formed. After 2,000 steps, the test sample does not show any clear changes. However, after 4,000 steps the test sample shows a clear stable pattern in the spatial distribution of species A.

The diffusion coefficients in E2 initially start with $D_A > D_B$. After around 500 gradient steps this relation is reversed and after 4000 steps $D_B$ is roughly twice as large as $D_A$. \cite{malevanets1997microscopic} note in their analysis of the system that the ratio $\sfrac{D_B}{D_A}$ needs to exceed a critical value for labyrinthine stable patterns to form. This agrees with the values found through optimization.

Finally, the results of experiment \textit{E3} are shown in Fig. \ref{res:fhn}C. Standing out immediately is that fitting the reaction rates takes significantly fewer gradient steps than the diffusion coefficients. After only 50 steps of gradient descent, the test sample shows the first patterns, albeit not appearing fully stable. Then after 500 steps, clear stable patterns are formed. Comparing the results of experiments E2 and E3, fitting the reaction rates appears to be significantly easier compared to fitting the diffusion coefficients. 

Overall, the results of the experiments show that it is possible to apply AD via approximate gradients in order to identify regions within the parameter space of growth \acp{sLM} that maintain given stable patterns. This was shown for two types of parameters; diffusion coefficients and reaction rates. While the results suggest that fitting diffusion coefficients is more challenging, the method was successfully in both cases.

\section{Discussion}

In this study, we investigated an alternative fitting approach for gradient-based optimization, using backpropagation of approximate gradients. To this end, four common types of fitting tasks pertaining to \ac{sLM} were identified. For each type, a series of experiments was performed to investigate the viability of the proposed method.

Taken together, our experimental results suggest that \ac{AD} via approximate gradients is a viable and versatile strategy to fit sLM parameters to data. As a proof of concept, we  apply \ac{AD} via approximate gradients to four simple models taken from different domains of research that require different fitting tasks. 
Further, we provide an easy-to-use codebase that can be referenced for further experiments. 

Similar fitting tasks as those shown here apply to some classes of Neural Networks. Specifically, optimization of the Malevanets-Kapral model is similar to optimization of generative Deep Diffusion models. These models are trained by iteratively injecting noise into (i.e., diffusing) an input pattern (e.g., an image) until the original pattern degenerates to white noise. Then, a parameterized network tries to  recover the original input pattern \cite{cao2022survey}. The challenge here is that the noise-generating process is unknown and needs to be estimated by the network. To achieve that, gradients need to be retained and backpropagated from the white noise patterns. For the Malevanets-Kapral model, we have a known reaction-diffusion process that destroys stable patterns over time. Instead of learning how to revert the noise process, our goal is to adjust the parameters of the process itself, such that the target image becomes a stable attractor of the system.

While gradient descent is often effective, it also introduces new hyperparameters. The learning rate, for example, can have a significant impact on convergence \cite{jacobs1988increased}. More sophisticated gradient descent algorithms introduce additional hyperparameters \cite{kingma2014adam}. For large neural networks with millions of parameters,  these additional hyperparameters are a comparatively small price to pay for efficient optimization. \Acp{sLM}, however, usually have relatively few parameters, and the number of hyperparameters could easily be larger. In this scenario, we have simply replaced the original parameter estimation problem with another one of possibly the same complexity. Further challenges of gradient-based optimization methods to consider are convergence to local optima and the problem of defining the right cost function.

In future research, we aim to apply the method to more complex \acp{sLM} such as the Cellular Potts Model for cells and tissues \cite{graner1992simulation}. For these more complex models, the performance of the method should be compared to parameter calibration via gradient-free calibration methods such as ABC. While these methods are not directly comparable in that they do not solve the same task (posterior inference compared to mode estimation), we could use for example the wall clock time of the methods to establish a comparison of efficacy for calibration.




\section{Acknowledgements}
Part of the research presented was performed for an ELLIS excellence fellowship. The ELLIS Excellence Fellowship was funded by Radboud AI, Radboud University.




\begin{thebibliography}{10}

\bibitem{alamoudi2022massively}
E~Alamoudi, J~Starru{\ss}, N~Bundgaard, R~Muller, F~Reck, F~Graw, L~Brusch,
  J~Hasenauer, and Y~Schalte.
\newblock Massively parallel likelihood-free parameter inference for biological
  multi-scale systems.
\newblock {\em Publication Series of the John von Neumann Institute for
  Computing}, 51:355--365, 2022.

\bibitem{alamoudi2023fitmulticell}
Emad Alamoudi, Yannik Sch\"{a}lte, Robert M\"{u}ller, J\"{o}rn Starru{\ss},
  Nils Bundgaard, Frederik Graw, Lutz Brusch, and Jan Hasenauer.
\newblock {FitMultiCell}: Simulating and parameterizing computational models of
  multi-scale and multi-cellular processes.
\newblock February 2023.

\bibitem{ali2023spatiotemporal}
Ishtiaq Ali and Maliha~Tehseen Saleem.
\newblock Spatiotemporal dynamics of reaction{\textendash}diffusion system and
  its application to turing pattern formation in a gray{\textendash}scott
  model.
\newblock {\em Mathematics}, 11(6):1459, March 2023.

\bibitem{andelfinger2021differentiable}
Philipp Andelfinger.
\newblock Differentiable agent-based simulation for gradient-guided
  simulation-based optimization.
\newblock In {\em Proceedings of the 2021 ACM SIGSIM Conference on Principles
  of Advanced Discrete Simulation}, SIGSIM-PADS '21, page 27–38, New York,
  NY, USA, 2021. Association for Computing Machinery.

\bibitem{andelfinger2023towards}
Philipp Andelfinger.
\newblock Towards differentiable agent-based simulation.
\newblock {\em ACM Trans. Model. Comput. Simul.}, 32(4), jan 2023.

\bibitem{arya2022automatic}
Gaurav Arya, Moritz Schauer, Frank Sch{\"a}fer, and Christopher~Vincent
  Rackauckas.
\newblock Automatic differentiation of programs with discrete randomness.
\newblock In Alice~H. Oh, Alekh Agarwal, Danielle Belgrave, and Kyunghyun Cho,
  editors, {\em Advances in Neural Information Processing Systems}, 2022.

\bibitem{beaumont2002approximate}
Mark~A Beaumont, Wenyang Zhang, and David~J Balding.
\newblock Approximate bayesian computation in population genetics.
\newblock {\em Genetics}, 162(4):2025--2035, December 2002.

\bibitem{bengio2013estimating}
Yoshua Bengio, Nicholas Léonard, and Aaron Courville.
\newblock Estimating or propagating gradients through stochastic neurons for
  conditional computation, 2013.

\bibitem{beyer7behavior}
Hans-Georg Beyer, Markus Olhofer, and Bernhard Sendhoff.
\newblock On the behavior of ($\mu$/$\mu${\'a}, $\lambda$)-es optimizing
  functions disturbed by generalized noise.
\newblock 01 2002.

\bibitem{boychuk2009stochastic}
Den Boychuk, W.~John Braun, Reg~J. Kulperger, Zinovi~L. Krougly, and David~A.
  Stanford.
\newblock A stochastic forest fire growth model.
\newblock {\em Environmental and Ecological Statistics}, 16(2):133--151, March
  2008.

\bibitem{cannon2022investigating}
Patrick Cannon, Daniel Ward, and Sebastian~M. Schmon.
\newblock Investigating the impact of model misspecification in neural
  simulation-based inference, 2022.

\bibitem{cao2022survey}
Hanqun Cao, Cheng Tan, Zhangyang Gao, Yilun Xu, Guangyong Chen, Pheng-Ann Heng,
  and Stan~Z. Li.
\newblock A survey on generative diffusion model, 2022.

\bibitem{carr2021estimating}
Michael~J. Carr, Matthew~J. Simpson, and Christopher Drovandi.
\newblock Estimating parameters of a stochastic cell invasion model with
  fluorescent cell cycle labelling using approximate bayesian computation.
\newblock {\em Journal of The Royal Society Interface}, 18(182):20210362,
  September 2021.

\bibitem{chang2021supporting}
Serina Chang, Mandy~L. Wilson, Bryan Lewis, Zakaria Mehrab, Komal~K. Dudakiya,
  Emma Pierson, Pang~Wei Koh, Jaline Gerardin, Beth Redbird, David Grusky,
  Madhav Marathe, and Jure Leskovec.
\newblock Supporting {COVID}-19 policy response with large-scale mobility-based
  modeling.
\newblock In {\em Proceedings of the 27th {ACM} {SIGKDD} Conference on
  Knowledge Discovery and Data Mining}. {ACM}, August 2021.

\bibitem{chopard2005cellular}
Bastien Chopard.
\newblock Cellular automata modeling of physical systems.
\newblock In {\em Encyclopedia of Complexity and Systems Science}, pages
  865--892. Springer New York, 2009.

\bibitem{chopra2022differentiable}
Ayush Chopra, Alexander Rodr\'{\i}guez, Jayakumar Subramanian, Arnau
  Quera-Bofarull, Balaji Krishnamurthy, B.~Aditya Prakash, and Ramesh Raskar.
\newblock Differentiable agent-based epidemiology.
\newblock In {\em Proceedings of the 2023 International Conference on
  Autonomous Agents and Multiagent Systems}, AAMAS '23, page 1848–1857,
  Richland, SC, 2023. International Foundation for Autonomous Agents and
  Multiagent Systems.

\bibitem{conlisk1976interactive}
John Conlisk.
\newblock Interactive markov chains.
\newblock {\em The Journal of Mathematical Sociology}, 4(2):157--185, January
  1976.

\bibitem{dab1991lattice}
David Dab, Jean-Pierre Boon, and Yue-Xian Li.
\newblock Lattice-gas automata for coupled reaction-diffusion equations.
\newblock {\em Physical Review Letters}, 66(19):2535--2538, May 1991.

\bibitem{durso2021hcv}
Karina Durso-Cain, Peter Kumberger, Yannik Sch\"{a}lte, Theresa Fink, Harel
  Dahari, Jan Hasenauer, Susan~L. Uprichard, and Frederik Graw.
\newblock {HCV} spread kinetics reveal varying contributions of transmission
  modes to infection dynamics.
\newblock {\em Viruses}, 13(7):1308, July 2021.

\bibitem{dyer2022black}
Joel Dyer, Patrick Cannon, J.~Doyne Farmer, and Sebastian Schmon.
\newblock Black-box bayesian inference for economic agent-based models, 2022.

\bibitem{fitzhugh1961impulses}
Richard FitzHugh.
\newblock Impulses and physiological states in theoretical models of nerve
  membrane.
\newblock {\em Biophysical Journal}, 1(6):445--466, July 1961.

\bibitem{frazier2020model}
David~T. Frazier, Christian~P. Robert, and Judith Rousseau.
\newblock Model misspecification in approximate bayesian computation:
  Consequences and diagnostics.
\newblock {\em Journal of the Royal Statistical Society Series B: Statistical
  Methodology}, 82(2):421--444, January 2020.

\bibitem{furth1920brownsche}
Reinhold F{\"u}rth.
\newblock Die brownsche bewegung bei ber{\"u}cksichtigung einer persistenz der
  bewegungsrichtung. mit anwendungen auf die bewegung lebender infusorien.
\newblock {\em Zeitschrift f{\"u}r Physik}, 2(3):244--256, June 1920.

\bibitem{graner1992simulation}
Fran{\c{c}}ois Graner and James~A. Glazier.
\newblock Simulation of biological cell sorting using a two-dimensional
  extended potts model.
\newblock {\em Physical Review Letters}, 69(13):2013--2016, September 1992.

\bibitem{hansen1995adaptation}
Nikolaus Hansen, Andreas Ostermeier, and Andreas Gawelczyk.
\newblock On the adaptation of arbitrary normal mutation distributions in
  evolution strategies: The generating set adaptation.
\newblock {\em Proceedings of the Sixth International Conference on Genetic
  Algorithms}, 01 1997.

\bibitem{haselwandter2008renormalization}
Christoph~A. Haselwandter and Dimitri~D. Vvedensky.
\newblock Renormalization of stochastic lattice models: Epitaxial surfaces.
\newblock {\em Physical Review E}, 77(6), June 2008.

\bibitem{ho2012perturbation}
Yu-Chi Ho and Xi-Ren Cao.
\newblock {\em Perturbation Analysis of Discrete Event Dynamic Systems}.
\newblock Springer {US}, 1991.

\bibitem{hu2021handling}
Zheng Hu, Jiaojiao Zhang, and Yun Ge.
\newblock Handling vanishing gradient problem using artificial derivative.
\newblock {\em {IEEE} Access}, 9:22371--22377, 2021.

\bibitem{huijben2022review}
Iris A.~M. Huijben, Wouter Kool, Max~B. Paulus, and Ruud J.~G. van Sloun.
\newblock A review of the gumbel-max trick and its extensions for discrete
  stochasticity in machine learning.
\newblock {\em {IEEE} Transactions on Pattern Analysis and Machine
  Intelligence}, 45(2):1353--1371, February 2023.

\bibitem{jacobs1988increased}
Robert~A. Jacobs.
\newblock Increased rates of convergence through learning rate adaptation.
\newblock {\em Neural Networks}, 1(4):295--307, January 1988.

\bibitem{jang2016categorical}
Eric Jang, Shixiang Gu, and Ben Poole.
\newblock Categorical reparameterization with gumbel-softmax, 2016.

\bibitem{kelly2023misspecification}
Ryan~P. Kelly, David~J. Nott, David~T. Frazier, David~J. Warne, and Chris
  Drovandi.
\newblock Misspecification-robust sequential neural likelihood, 2023.

\bibitem{kingma2014adam}
Diederik~P. Kingma and Jimmy Ba.
\newblock Adam: A method for stochastic optimization, 2014.

\bibitem{kingma2013auto}
Diederik~P Kingma and Max Welling.
\newblock Auto-encoding variational bayes, 2013.

\bibitem{lengyel1992chemical}
I~Lengyel and I~R Epstein.
\newblock A chemical approach to designing turing patterns in
  reaction-diffusion systems.
\newblock {\em Proceedings of the National Academy of Sciences},
  89(9):3977--3979, May 1992.

\bibitem{li2014efficient}
Mu~Li, Tong Zhang, Yuqiang Chen, and Alexander~J. Smola.
\newblock Efficient mini-batch training for stochastic optimization.
\newblock In {\em Proceedings of the 20th ACM SIGKDD International Conference
  on Knowledge Discovery and Data Mining}, KDD '14, page 661–670, New York,
  NY, USA, 2014. Association for Computing Machinery.

\bibitem{li2015turing}
Shanbing Li, Jianhua Wu, and Yaying Dong.
\newblock Turing patterns in a reaction{\textendash}diffusion model with the
  degn{\textendash}harrison reaction scheme.
\newblock {\em Journal of Differential Equations}, 259(5):1990--2029, September
  2015.

\bibitem{long2018foreground}
Long~Ang Lim and Hacer~Yalim Keles.
\newblock Foreground segmentation using convolutional neural networks for
  multiscale feature encoding.
\newblock {\em Pattern Recognition Letters}, 112:256--262, September 2018.

\bibitem{lim2020learning}
Long~Ang Lim and Hacer~Yalim Keles.
\newblock Learning multi-scale features for foreground segmentation.
\newblock {\em Pattern Analysis and Applications}, 23(3):1369--1380, August
  2019.

\bibitem{ljosa2012annotated}
Vebjorn Ljosa, Katherine~L Sokolnicki, and Anne~E Carpenter.
\newblock Annotated high-throughput microscopy image sets for validation.
\newblock {\em Nature Methods}, 9(7):637--637, June 2012.

\bibitem{lueckmann2021benchmarking}
Jan-Matthis Lueckmann, Jan Boelts, David Greenberg, Pedro Goncalves, and Jakob
  Macke.
\newblock Benchmarking simulation-based inference.
\newblock In Arindam Banerjee and Kenji Fukumizu, editors, {\em Proceedings of
  The 24th International Conference on Artificial Intelligence and Statistics},
  volume 130 of {\em Proceedings of Machine Learning Research}, pages 343--351.
  PMLR, 13--15 Apr 2021.

\bibitem{maddison2014sampling}
Chris~J. Maddison, Daniel Tarlow, and Tom Minka.
\newblock A* sampling, 2014.

\bibitem{malevanets1997microscopic}
Anatoly Malevanets and Raymond Kapral.
\newblock Microscopic model for {FitzHugh}-nagumo dynamics.
\newblock {\em Physical Review E}, 55(5):5657--5670, May 1997.

\bibitem{mengersen2013bayesian}
Kerrie~L. Mengersen, Pierre Pudlo, and Christian~P. Robert.
\newblock Bayesian computation via empirical likelihood.
\newblock {\em Proceedings of the National Academy of Sciences},
  110(4):1321--1326, January 2013.

\bibitem{merritt2013dynamics}
David Merritt.
\newblock {\em Dynamics and Evolution of Galactic Nuclei}.
\newblock Princeton University Press, July 2013.

\bibitem{mordvintsev2020growing}
Alexander Mordvintsev, Ettore Randazzo, Eyvind Niklasson, and Michael Levin.
\newblock Growing neural cellular automata.
\newblock {\em Distill}, 5(2), February 2020.

\bibitem{neftci2019surrogate}
Emre~O. Neftci, Hesham Mostafa, and Friedemann Zenke.
\newblock Surrogate gradient learning in spiking neural networks: Bringing the
  power of gradient-based optimization to spiking neural networks.
\newblock {\em {IEEE} Signal Processing Magazine}, 36(6):51--63, November 2019.

\bibitem{paszke2017automatic}
Adam Paszke, Sam Gross, Soumith Chintala, Gregory Chanan, Edward Yang, Zachary
  DeVito, Zeming Lin, Alban Desmaison, Luca Antiga, and Adam Lerer.
\newblock Automatic differentiation in pytorch.
\newblock 2017.

\bibitem{qin2022highly}
Xuebin Qin, Hang Dai, Xiaobin Hu, Deng-Ping Fan, Ling Shao, and Luc~Van Gool.
\newblock Highly accurate dichotomous image segmentation.
\newblock In {\em Lecture Notes in Computer Science}, pages 38--56. Springer
  Nature Switzerland, 2022.

\bibitem{quera2023don}
Arnau Quera-Bofarull, Ayush Chopra, Joseph Aylett-Bullock, Carolina
  Cuesta-Lazaro, Anisoara Calinescu, Ramesh Raskar, and Michael Wooldridge.
\newblock Don't simulate twice: One-shot sensitivity analyses via automatic
  differentiation.
\newblock In {\em Proceedings of the 2023 International Conference on
  Autonomous Agents and Multiagent Systems}, AAMAS '23, page 1867–1876,
  Richland, SC, 2023. International Foundation for Autonomous Agents and
  Multiagent Systems.

\bibitem{quera2023some}
Arnau Quera-Bofarull, Joel Dyer, Anisoara Calinescu, and Michael Wooldridge.
\newblock Some challenges of calibrating differentiable agent-based models,
  2023.

\bibitem{romero2021public}
Santiago Romero-Brufau, Ayush Chopra, Alex~J Ryu, Esma Gel, Ramesh Raskar,
  Walter Kremers, Karen~S Anderson, Jayakumar Subramanian, Balaji
  Krishnamurthy, Abhishek Singh, Kalyan Pasupathy, Yue Dong, John~C O'Horo,
  Walter~R Wilson, Oscar Mitchell, and Thomas~C Kingsley.
\newblock Public health impact of delaying second dose of {BNT}162b2 or
  {mRNA}-1273 covid-19 vaccine: simulation agent based modeling study.
\newblock {\em {BMJ}}, page n1087, May 2021.

\bibitem{rothman2004lattice}
Daniel~H. Rothman and Stiphane Zaleski.
\newblock {\em Lattice-Gas Cellular Automata}.
\newblock Cambridge University Press, August 1997.

\bibitem{scarpino2019predictability}
Samuel~V. Scarpino and Giovanni Petri.
\newblock On the predictability of infectious disease outbreaks.
\newblock {\em Nature Communications}, 10(1), February 2019.

\bibitem{scholes2019comprehensive}
Natalie~S. Scholes, David Schnoerr, Mark Isalan, and Michael~P.H. Stumpf.
\newblock A comprehensive network atlas reveals that turing patterns are common
  but not robust.
\newblock {\em Cell Systems}, 9(3):243--257.e4, September 2019.

\bibitem{scianna2021cellular}
Marco Scianna and Luigi Preziosi.
\newblock A cellular potts model for analyzing cell migration across
  constraining pillar arrays.
\newblock {\em Axioms}, 10(1):32, March 2021.

\bibitem{shi2021numerical}
Hao-Jun~Michael Shi, Melody~Qiming Xuan, Figen Oztoprak, and Jorge Nocedal.
\newblock On the numerical performance of derivative-free optimization methods
  based on finite-difference approximations, 2021.

\bibitem{sisson2018handbook}
S.~A. Sisson, Y.~Fan, and M.~A. Beaumont, editors.
\newblock {\em Handbook of Approximate Bayesian Computation}.
\newblock Chapman and Hall/{CRC}, September 2018.

\bibitem{sisson2007sequential}
S.~A. Sisson, Y.~Fan, and Mark~M. Tanaka.
\newblock Sequential monte carlo without likelihoods.
\newblock {\em Proceedings of the National Academy of Sciences},
  104(6):1760--1765, February 2007.

\bibitem{stevens2020deep}
Eli Stevens, Luca Antiga, and Thomas Viehmann.
\newblock {\em Deep learning with PyTorch}.
\newblock Manning Publications, 2020.

\bibitem{storath2015joint}
Martin Storath, Andreas Weinmann, J\"{u}rgen Frikel, and Michael Unser.
\newblock Joint image reconstruction and segmentation using the potts model.
\newblock {\em Inverse Problems}, 31(2):025003, January 2015.

\bibitem{sumata2013comparison}
H.~Sumata, F.~Kauker, R.~Gerdes, C.~K\"{o}berle, and M.~Karcher.
\newblock A comparison between gradient descent and stochastic approaches for
  parameter optimization of a sea ice model.
\newblock {\em Ocean Science}, 9(4):609--630, July 2013.

\bibitem{szabo2013cellular}
Andr{\'{a}}s Szab{\'{o}} and Roeland M.~H. Merks.
\newblock Cellular potts modeling of tumor growth, tumor invasion, and tumor
  evolution.
\newblock {\em Frontiers in Oncology}, 3, 2013.

\bibitem{tang2022relaxation}
Jianxiong Tang, Jian-Huang Lai, Wei-Shi Zheng, Lingxiao Yang, and Xiaohua Xie.
\newblock Relaxation {LIF}: A gradient-based spiking neuron for direct training
  deep spiking neural networks.
\newblock {\em Neurocomputing}, 501:499--513, August 2022.

\bibitem{tsekov2013brownian}
Roumen Tsekov and Marga~C. Lensen.
\newblock Brownian motion and the temperament of living cells.
\newblock {\em Chinese Physics Letters}, 30(7):070501, July 2013.

\bibitem{turing1990chemical}
Alan~Mathison Turing.
\newblock The chemical basis of morphogenesis.
\newblock {\em Philosophical Transactions of the Royal Society of London.
  Series B, Biological Sciences}, 237(641):37--72, August 1952.

\bibitem{vittadello2021turing}
Sean~T. Vittadello, Thomas Leyshon, David Schnoerr, and Michael P.~H. Stumpf.
\newblock Turing pattern design principles and their robustness.
\newblock {\em Philosophical Transactions of the Royal Society A: Mathematical,
  Physical and Engineering Sciences}, 379(2213), November 2021.

\bibitem{wang2011numerical}
Weiming Wang, Yezhi Lin, Feng Yang, Lei Zhang, and Yongji Tan.
\newblock Numerical study of pattern formation in an extended
  gray{\textendash}scott model.
\newblock {\em Communications in Nonlinear Science and Numerical Simulation},
  16(4):2016--2026, April 2011.

\bibitem{wang2022calibration}
Xiaoyu Wang, Adrianne~L. Jenner, Robert Salomone, David~J. Warne, and
  Christopher Drovandi.
\newblock Calibration of agent based models for monophasic and biphasic tumour
  growth using approximate bayesian computation.
\newblock September 2022.

\bibitem{wang2017interactive}
Yi~Wang, Zhiming Luo, and Pierre-Marc Jodoin.
\newblock Interactive deep learning method for segmenting moving objects.
\newblock {\em Pattern Recognition Letters}, 96:66--75, September 2017.

\bibitem{ward2022robust}
Daniel Ward, Patrick Cannon, Mark Beaumont, Matteo Fasiolo, and Sebastian~M
  Schmon.
\newblock Robust neural posterior estimation and statistical model criticism,
  2022.

\bibitem{wortel2021celltrackr}
Inge~M.N. Wortel, Annie~Y. Liu, Katharina Dannenberg, Jeffrey~C. Berry, Mark~J.
  Miller, and Johannes Textor.
\newblock {CelltrackR}: An r package for fast and flexible analysis of immune
  cell migration data.
\newblock {\em {ImmunoInformatics}}, 1-2:100003, October 2021.

\bibitem{wortel2021local}
Inge~M.N. Wortel, Ioana Niculescu, P.~Martijn Kolijn, Nir~S. Gov, Rob~J.
  de~Boer, and Johannes Textor.
\newblock Local actin dynamics couple speed and persistence in a cellular potts
  model of cell migration.
\newblock {\em Biophysical Journal}, 120(13):2609--2622, July 2021.

\end{thebibliography}


\end{document}